\documentclass[aps,pra,showpacs,floatfix,preprint]{revtex4}

\usepackage{bm}
\usepackage{graphicx}
\usepackage{amsmath}
\usepackage{amssymb}

\newcommand{\eps}{\varepsilon}
\newcommand{\la}{\lambda}

\newcommand{\om}{\omega}

\begin{document}

\title{Dynamics of Bose-Einstein condensates in cigar-shaped traps}

\author{A.M. Kamchatnov$^{1}$\footnote{Electronic Address: \tt kamch@isan.troitsk.ru}}

\author{V.S. Shchesnovich$^{2}$\footnote{Present address: Departamento de F\'{\i}sica - Universidade
Federal de Alagoas, Macei\'o AL 57072-970, Brazil}}

\affiliation{
$^1$Institute of Spectroscopy, Russian Academy of Sciences, Troitsk
142190, Moscow Region, Russia\\
$^2$Instituto de F\'{\i}sica Te\'{o}rica, Universidade Estadual
Paulista-UNESP, Rua Pamplona 145, 01405-900 S\~{a}o Paulo, Brazil }

\date{\today}


\begin{abstract}

Gross-Pitaevskii equation for Bose-Einstein condensate confined in
elongated cigar-shaped trap is reduced to an effective system of nonlinear
equations depending on only one space coordinate along the trap axis. The
radial distribution of the condensate density and its radial velocity are
approximated by Gaussian functions with real and imaginary exponents,
respectively, with parameters depending on the axial coordinate and time.
The effective one-dimensional system  is applied to description of the
ground state of the condensate, to dark and bright solitons, to the
sound and radial compression waves propagating in a dense condensate,
and to weakly nonlinear waves in repulsive condensate. In
the low density limit our results reproduce the known formulae. In the high
density case our description of solitons goes beyond the standard approach
based on the nonlinear Schr\"odinger equation. The dispersion relations for
the sound and radial compression waves are obtained in a wide region of
values of the condensate density. Korteweg-de Vries equation for weakly
nonlinear waves is derived and existence of bright solitons on a
constant background is predicted for the dense enough condensate with
repulsive interaction between the atoms.

\end{abstract}

\maketitle

\section{Introduction}

Formation of Bose-Einstein condensate (BEC) in dilute gases
\cite{bec1,bec2,bec3} has created a new field of nonlinear physics with new
specific problems. Among new features of BEC in gases are the spatial
inhomogeneity created by the confining potential,  interplay of coherent
and nonlinear phenomena, multi-species condensation, etc. Many of such new
problems have already been addressed in current literature. In particular,
the condensate oscillations near the center of the trap and its free
expansion after switching off the trap potential was considered in Refs.
\cite{CastinDum96,Kagan96-1,Kagan96-2,Dalfovo97,SZ02,Plaja02,Das02-1,kamch04}.
Creation of strongly elongated (``cigar-shaped'') traps has made possible
to generate dark \cite{burger,denshlag} and bright
\cite{khaykovich,strecker} solitons in BEC. However, theoretical
description of the solitons proves to be not a simple problem. Indeed, the
standard reduction of the Gross-Pitaevskii (GP) equation to the
one-dimensional (1D) nonlinear Schr\"odinger (NLS) equation is justified
only under the condition that axial size of the solution (along the trap
axis) is much greater than its radial size. However, the actual solitons
observed in experiment \cite{khaykovich,strecker} have comparable axial and
radial sizes. An attempt to go beyond the NLS approximation was made in
Ref.~\cite{salasnich} under the assumption that variation of the field
variables in the axial direction takes place on much longer scale than that
in the radial direction. Therefore the equation with a ``non-polynomial
interaction'' obtained in Ref.~\cite{salasnich} describes the ground
state of BEC with repulsive interaction when the above assumption is
correct but its application to ``short'' dark or bright solitons is not
well justified.

In this paper we consider dynamics of BEC in the cigar-shaped trap by means
of variational approach without assumption  that the field variables change
slowly along the trap axis. The radial distribution of BEC density and its
radial velocity are approximated by real and imaginary Gaussian functions,
respectively, with parameters depending on the axial coordinate $z$. This
approach allows us to reduce the three-dimensional (3D) BEC Lagrangian to
an effective 1D one which depends on the axial wave function $\Psi(z,t)$,
the mean condensate radius $b(z,t)$, and the radial velocity potential
$\alpha(z,t)$. The resulting system of equations for the above three
variables, though being rather complicated, has an advantage of dependence
on only one spatial coordinate $z$. It contains as limiting cases the NLS
and the ``non-polynomial interaction'' approximations. The derived
effective system allows one to investigate bright and dark solitons in a
wide domain of their existence. In particular, the bright solitons can be
considered right up to the collapse instability threshold. We also study
the linear waves propagating along the condensate with repulsive
interaction between the atoms. There are two types of waves: the low
frequency sound waves and the high frequency radial compression wave. The
dispersion relations for the linear waves, applicable to the high density
BEC, are obtained in the long wavelength limit. In the next approximation
of ``small but finite'' amplitude of the ground state perturbation we
obtain the Korteweg-de Vries (KdV) equation for weakly dispersive and
nonlinear waves propagating along the condensate. An interesting feature of
this equation is that in a high density BEC the coefficient in the
dispersion term changes sign. An immediate consequence is that fast
solitons of small amplitude, propagating in a repulsive BEC, are bright
rather than dark, contrary to what is usually assumed by analogy between
the GP and 1D NLS equations.

\section{General quasi-1D equations for BEC evolution}

The starting point for our consideration is the GP equation for the wave
function $\psi$ of the condensate,
\begin{equation}\label{GP1}
    i\hbar\psi_t=-\frac{\hbar^2}{2m}\triangle\psi+\frac12m\omega_\bot^2
    r^2\psi+V(z)\psi+g|\psi|^2\psi,
\end{equation}
where $r=\sqrt{x^2+y^2}$ is the radial coordinate, $g=4\pi\hbar^2a_s/m$ is
the effective nonlinear coupling constant, $a_s$ denotes  the $s$-wave
scattering length, $m$ is the atomic mass, $\omega_\bot$ is the frequency
of atomic oscillations in the radial direction, and $V(z)$ is the trap
potential in the axial direction of a cigar-shaped trap. The condensate
wave function $\psi$ is normalized to the number $N$ of atoms,
\begin{equation}\label{norm3D}
    \int|\psi|^2d \mathbf{r}=N.
\end{equation}
The GP equation admits the least action principle  formulation with the
action functional
\begin{equation}\label{action3D}
    S=\int Ldt,\quad L=\int\mathcal{L}\cdot 2\pi rdrdz,
\end{equation}
written under assumption that the condensate motion is axially symmetric,
where the Lagrangian density is given by
\begin{equation}\label{lagr3D}
    \mathcal{L}=\frac{i\hbar}2\left(\psi\psi_t^*-\psi^*\psi_t\right)+
    \frac{\hbar^2}{2m}|\nabla\psi|^2+\frac12m\omega_\bot r^2|\psi|^2+
    V(z)|\psi|^2+\frac{g}2|\psi|^4.
\end{equation}

It is well known (see, e.g. \cite{perez}) that if the linear (1D) density
$n_1=\int_0^\infty|\psi|^22\pi r dr$ of the condensate is small enough,
\begin{equation}\label{c1}
    |a_s|n_1\ll 1,
\end{equation}
then the 3D GP equation can be reduced to the 1D NLS equation
\begin{equation}\label{c2}
    i\hbar\Psi_t=-\frac{\hbar^2}{2m}\Psi_{zz}+V(z)\Psi+g_{1D}|\Psi|^2\Psi,
\end{equation}
where
\begin{equation}\label{c3}
    g_{1D}=\frac{g}{2\pi a_\bot^2}=\frac{2\hbar^2a_s}{ma_\bot^2},\quad
    a_\bot=\sqrt{\frac{\hbar}{m\om_\bot}}.
\end{equation}
This reduction can be performed, for example, by variational method, when
under supposition (\ref{c1}) the transversal degrees of freedom are
``frozen'' and transverse motion is reduced to the ground state
oscillations in the radial direction. As a result, the condensate wave
function $\psi$ can be factorized as
\begin{equation}\label{c4}
    \psi=\frac1{\sqrt{\pi}a_\bot}\exp\left(-\frac{r^2}{2a_\bot^2}\right)
    \Psi(z,t).
\end{equation}
Its substitution into (\ref{action3D}), (\ref{lagr3D}) and integration over
radial coordinate yield an effective 1D Lagrangian and the corresponding
Lagrange equation for $\Psi(z,t)$ is Eq.~(\ref{c2}) (see, e.g.
\cite{kamch04}). Here we want to generalize this procedure to the case when
$|a_s|n_1\sim1$ and the radial motion of the condensate must be taken into
account.

Assuming quasi-1D geometry, we approximate the wave function as
\begin{equation}\label{wave1}
    \psi=\frac1{\sqrt{\pi}b(z,t)}\exp\left(-\frac{r^2}{2b^2(z,t)}\right)
    \exp\left(\frac{i}2\alpha(z,t)r^2\right)\Psi(z,t).
\end{equation}
Here $b(z,t)$ characterizes the local radius of the condensate and
$\alpha(z,t)r$ gives the local radial velocity.  Substitution of
(\ref{wave1}) into (\ref{lagr3D}) and the result into (\ref{action3D}) with
integration over the radial coordinate yield an effective 1D Lagrangian
\begin{equation}\label{1D-lagr-1}
    \begin{split}
    \mathcal{L}_{1D}=\frac{i\hbar}2\left(\Psi\Psi_t^*-\Psi^*\Psi_t\right)+
    \frac{\hbar^2}{2m}|\Psi_z|^2+\frac12m\omega_\bot^2b^2|\Psi|^2+
    V(z)|\Psi|^2+\frac{g}{4\pi b^2}|\Psi|^4\\
    +\frac\hbar2\alpha_tb^2|\Psi|^2+\frac{\hbar^2}{2m}\left[\left(
    \frac1{b^{2}}+\alpha^2b^2+\frac{b_z^2}{b^{2}}+\frac12\alpha_z^2b^4\right)|\Psi|^2
    +\frac{i}2\alpha_zb^2\left(\Psi\Psi_z^*-\Psi^*\Psi_z\right)\right].
    \end{split}
\end{equation}
To simplify the notation, let us introduce dimensionless variables:
\begin{equation}\label{dim}
    \tau=\omega_\bot t,\quad z=a_\bot\zeta,\quad \Psi=\frac{u}{\sqrt{a_\bot}},
    \quad b=a_\bot w,\quad \alpha=\frac\beta{a_\bot^2},\quad
    G=\frac{a_s}{a_\bot}.
\end{equation}
In the dimensionless variables the 1D Lagrangian becomes
\begin{equation}\label{1D-lagr-2}
\begin{split}
    \mathcal{L}_{1D}&=\frac{\hbar^2}{ma_\bot^3}\Bigg[\frac{i}2(uu_\tau^*-u^*u_\tau)
    +\frac12|u_\zeta|^2+\frac12w^2|u|^2+V(\zeta)|u|^2+G\frac{|u|^4}{w^2}\\ &+
    \frac12\beta_\tau w^2|u|^2+\frac12(w^{-2}+\beta^2w^2+w_\zeta^2w^{-2}+
    \frac12\beta_\zeta^2w^4)|u|^2+\frac{i}4\beta_\zeta w^2(uu_\zeta^*-
    u^*u_\zeta)\Bigg].
    \end{split}
\end{equation}
Thus, in the above  reduction, the evolution of BEC is described by the
complex longitudinal wave function $u(\zeta,\tau)$ and two real functions,
$w(\zeta,\tau)$ and $\beta(\zeta,\tau)$, corresponding to the local mean
radius of the condensate and its radial velocity, respectively. Equations,
governing the evolution of these variables are to be obtained from the
action principle
\begin{equation}\label{action-1D}
    S=\int Ldt=\mathrm{min},\quad L=\int\mathcal{L}_{1D}d\zeta
\end{equation}
with the effective Lagrangian density given by Eq.~(\ref{1D-lagr-2}). We
have:
\begin{equation}\label{eq-u}
\begin{split}
    iu_\tau &+\frac12u_{\zeta\zeta}-V(\zeta)u-3G\frac{|u|^2u}{w^2}\\
    &-\frac12\left[\frac2{w^2}+\frac{w_{\zeta\zeta}}w+\left(\frac{w_\zeta}w-
    \frac{i}2\beta_\zeta w^2\right)\frac{(|u|^2)_\zeta}{|u|^2}-
    \frac12\beta_\zeta^2w^4-\frac{i}2\beta_{\zeta\zeta}w^2-i\beta_\zeta
    ww_\zeta\right]=0,
    \end{split}
\end{equation}
\begin{equation}\label{eq-w}
    (w^2|u|^2)_\tau+\left[\tfrac{i}2w^2(uu_\zeta^*-u^*u_\zeta)
    +w^4|u|^2\beta_\zeta\right]=2\beta w^2|u|^2,
\end{equation}
\begin{equation}\label{eq-beta}
    \beta_\tau=\frac{w_{\zeta\zeta}}{w^3}-\frac{w_\zeta^2}{w^4}+
    \frac{w_\zeta}{w^3}\frac{(|u|^2)_\zeta}{|u|^2}+\frac1{w^4}-1
    -\beta^2-\beta_\zeta^2w^2+\frac{i(u_\zeta u^*-uu_\zeta^*)}{2|u|^2}
    \beta_\zeta+2G\frac{|u|^2}{w^4},
\end{equation}
where the longitudinal wave function is normalized as follows
\begin{equation}\label{norm1D}
    \int|u|^2d\zeta=N.
\end{equation}
Thus, we have transformed the 3D GP equation to a quasi-1D form for the
case of elongated cigar-shaped geometry when the radial distributions of
the condensate  density and its radial velocity can be approximated by
simple Gaussian functions. The variables in
Eqs.~(\ref{eq-u})--(\ref{eq-beta}) depend only on one spatial coordinate---%
a strong advantage in numerical simulations. Besides that,
system (\ref{eq-u})--(\ref{eq-beta}) can be analyzed analytically in some
important limiting cases (see below).

For some applications it is convenient to cast
Eqs.~(\ref{eq-u})--(\ref{eq-beta}) into a hydrodynamic-like form by means
of the  substitution
\begin{equation}\label{madelung}
    u=\sqrt{\rho}\,\exp\left(i\int^\zeta v(\zeta',\tau)d\zeta'\right),
\end{equation}
where $\rho(\zeta,\tau)$ is the density of BEC and $v(\zeta,\tau)$ denotes
its axial velocity. As a result, we obtain the hydrodynamic system:
\begin{equation}\label{cont}
    \rho_\tau+\left(\rho(v+\tfrac12w^2\beta_\zeta)\right)_\zeta=0,
\end{equation}
\begin{equation}\label{euler}
    v_\tau+vv_\zeta+V'(\zeta)+3G\left(\frac\rho{w^2}\right)_\zeta+
    \frac12\left(\frac{\rho_\zeta^2}{4\rho^2}-\frac{\rho_{\zeta\zeta}}{2\rho}
    +\frac2{w^2}+\frac{w_{\zeta\zeta}}w+\frac{w_\zeta\rho_\zeta}{w\rho}
    -\frac12\beta_\zeta^2w^4\right)_\zeta=0,
\end{equation}
\begin{equation}\label{beta-2}
    \beta_\tau=\frac{w_{\zeta\zeta}}{w^3}-\frac{w_\zeta^2}{w^4}+
    \frac{w_\zeta\rho_\zeta}{w^3\rho}+\frac1{w^4}-1
    -\beta^2-\beta_\zeta^2w^2+ v\beta_\zeta+2G\frac{\rho}{w^4},
\end{equation}
\begin{equation}\label{cont-2}
    (w^2\rho)_\tau+(w^2\rho v+w^4\rho\beta_\zeta)_\zeta=2\beta w^2\rho.
\end{equation}
In the following sections  we apply the general quasi-1D equations to some
important particular cases.

\section{Stationary solutions}

In a stationary state all velocities are equal to zero ($v=0,\,\beta=0$),
density $\rho$ does not depend on $\tau$ and the condensate wave function
$u(\zeta,\tau)$ depends on $\tau$ only through the phase factor
$\exp(-i\mu\tau)$, where $\mu$ is dimensionless chemical potential. Setting
$\beta=0$ and introducing the stationary variables $U(\zeta)$ and
$\sigma(\zeta)$ by the formulae
\begin{equation}\label{U-W}
    u(\zeta,\tau)=e^{-i\mu\tau}U(\zeta),\quad w^2(\zeta,\tau)=1/\sigma(\zeta),
\end{equation}
after simple algebra we obtain from Eqs. (\ref{eq-u})--(\ref{eq-beta})
the following system
\begin{equation}\label{eq-U2}
    U_{\zeta\zeta}+\left(2\mu-2V(\zeta)-\sigma-\frac1\sigma
    -\frac{\sigma_\zeta^2}{4\sigma^2}\right)U-4G\sigma{U^3}=0,
\end{equation}
\begin{equation}\label{eq-sigma}
    \frac12\left(\frac{\sigma_\zeta U^2}{\sigma}\right)_\zeta
    -\left(\sigma-\frac1\sigma\right)U^2-2G\sigma U^4=0.
\end{equation}
System (\ref{eq-U2})--(\ref{eq-sigma}) has the  energy functional
\begin{equation}\label{en-3}
    \mathcal{E}=\int\left[\frac12U_\zeta^2+V(\zeta)U^2+G\sigma U^4+
    \frac12\left(\sigma+\frac1\sigma+\frac{\sigma_\zeta^2}{4\sigma^2}
    \right)U^2-\mu U^2\right]d\zeta.
\end{equation}
Here integrand can be considered as a Lagrangian with the spatial variable
$\zeta$ playing the role of time. It is worth noticing that if we introduce
instead of $\sigma$ an angle $\theta$ by the relation
\begin{equation}\label{theta}
    \sigma=\exp(2\theta),
\end{equation}
then the energy functional takes the form
\begin{equation}\label{ham-1}
    \mathcal{E}=\int\left[\frac12\left(U_\zeta^2+U^2\theta_\zeta^2\right)
    +(V(\zeta)+\cosh2\theta-\mu))U^2+Ge^{2\theta}U^4\right]d\zeta
\end{equation}
which is formally an action functional of a particle moving on the plane,
where the couple $(U,\theta)$ assumes the role of the polar coordinates and
the external potential is
\begin{equation}\label{pot}
    \Phi(U,\theta)=(\mu-V(\zeta)-\cosh2\theta)U^2
    -Ge^{2\theta}U^4.
\end{equation}

\subsection{Small-density limit}
First of all let us consider the limit of a low density BEC:
\begin{equation}\label{low-dens-crit}
    |G|U^2\ll 1.
\end{equation}
In the dimensional variables,  with account of the estimate
$U^2\sim|\Psi|^2a_\bot\sim n_1a_\bot$ (where $n_1$ is the longitudinal
density of the condensate),  condition (\ref{low-dens-crit}) coincides with
(\ref{c1}). Since the characteristic length in the axial direction cannot
be less than unity (i.e. $a_\bot$ in the dimensional units),
we find in this limit from (\ref{eq-sigma})
\begin{equation}\label{sigma-low}
    \sigma=1+\sigma_1,\qquad |\sigma_1|\sim |G|U^2\ll 1.
\end{equation}
Taking estimate (\ref{sigma-low}) into account, we reduce Eq. (\ref{eq-U2})
to the standard (stationary) 1D NLS equation
\begin{equation}\label{NLS}
    U_{\zeta\zeta}+2(\mu-1-V(\zeta))U-4GU^3=0.
\end{equation}
This equation is studied very well and we shall not consider it any
further.

\subsection{Ground state of BEC with repulsive interaction
 ($G>0$)}

Another important limiting case corresponds to very slow dependence of BEC
parameters on the axial coordinate $\zeta$ so that one can neglect the
derivatives in Eqs.~(\ref{eq-U2}), (\ref{eq-sigma}). If we denote by $Z$
the characteristic axial length, then the condition for neglecting the
spatial derivatives can be written as
\begin{equation}\label{cond2}
    Z^{-2}\ll G\sigma U^2,
\end{equation}
or in the dimensional units, with $Z\sim l/a_\bot,$ $G=a_s/a_\bot,$ $U^2\sim
n_1a_\bot\sim Na_\bot/l,$ $\sigma\sim1$, as
\begin{equation}\label{cond3}
    \frac{a_\bot^2}{l^2}\ll\frac{aN}l.
\end{equation}
If  condition (\ref{cond3}) is satisfied, then Eq.~(\ref{eq-sigma}) gives
the relation
\begin{equation}\label{alg-rel}
    \sigma=(1+2GU^2)^{-1/2}
\end{equation}
equivalent to one obtained in Ref.~\cite{salasnich}. Substitution of
(\ref{alg-rel}) into (\ref{eq-U2}) and dropping off the derivatives relates
the field $U$ to the external trap:
\begin{equation}\label{mu-stat}
    \mu-V(\zeta)=\frac{1+3GU^2}{\sqrt{1+2GU^2}}.
\end{equation}
Solution of this equation with respect to $U^2$ yields the condensate
density as a function of the axial coordinate
\begin{equation}\label{TF}
    \rho(\zeta)=U^2=\frac1{2G}\left\{\frac19\left[\mu-V(\zeta)
    +\sqrt{(\mu-V(\zeta))^2
    +3}\right]^2-1\right\}.
\end{equation}
This formula corresponds to well-known Thomas-Fermi approximation for
condensate's density distribution. The density $\rho$ vanishes at
$\zeta=Z_m$ where value of $Z_m$ is determined by the equation
\begin{equation}\label{Zm}
    \mu-V(Z_m)=1
\end{equation}
which gives the relationship between $\mu$ and $Z_m$. The complementary
relation between $Z_m$ and  $\mu$ is given by the condition
\begin{equation}\label{norm3}
    \int\rho(\zeta)d\zeta=N,
\end{equation}
where the integration is taken over region where $\rho\geq0$. These two
relations allow us to express $\mu$ and $Z_m$ as functions of the number of
atoms $N$.

In the low density limit, i.e. $GU^2\ll1$, Eq.~(\ref{TF}) reduces to the
usual formula
\begin{equation}\label{TF2}
    \rho\cong\frac1{2G}(\mu'-V(\zeta)),\quad \mu'=\mu-1,
\end{equation}
which also follows from the 1D NLS equation (\ref{NLS}) of the preceding
subsection.

For the parabolic potential
\begin{equation}\label{parabolic}
    V(\zeta)=\frac12\la^2\zeta^2,\quad \la=\frac{\om_z}{\om_\bot},
\end{equation}
from Eq.~(\ref{Zm}) we obtain  ${\mu'}=\tfrac12\la^2\zeta_m^2$ so that
Eq.~(\ref{TF2}) with account of (\ref{norm3}) yields the well-known
expression for density distribution in the Thomas-Fermi approximation:
\begin{equation}\label{TF3}
    \rho(\zeta)=U^2=\frac{\la^2\zeta_m^2}{4G}\left(1-\frac{\zeta^2}{\zeta_m^2}
    \right),\quad \zeta_m=(3GN\la^{-2})^{1/3}.
\end{equation}

\subsection{Dark solitons in BEC with repulsive interaction}

It is well known that the NLS equation (\ref{NLS}) with $V(\zeta)=0$ and
$G>0$ has a dark soliton solution
\begin{equation}\label{dark-NLS}
    U(\zeta)=U_0\tanh(U_0\sqrt{2G}\,\zeta)
\end{equation}
under condition (\ref{low-dens-crit}), that is ${\mu'}\sim GU_0^2\ll 1$.
The soliton width is much greater than unity, $1/(U_0\sqrt{G})\gg 1$, and
it must be much less than the longitudinal size $\zeta_m$ of the condensate
estimated in the preceding subsection. For BEC confined by a parabolic trap
we have $U_0^3\sim N/\zeta_m$ so that the formulated condition reads
\begin{equation}\label{dark-sol-cond}
    \frac1{\zeta_m^2}\ll\frac{GN}{\zeta_m}\ll 1,
\end{equation}
where the left inequality actually coincides with condition  (\ref{cond2})
in the small density limit and the right inequality is equivalent to
(\ref{low-dens-crit}). Substitution of $\zeta_m$ from Eq.~(\ref{TF3})
yields the applicability condition of the soliton solution (\ref{dark-NLS})
in the form
\begin{equation}\label{dark-cond2}
    \sqrt{\la}\ll GN\ll \la^{-1}.
\end{equation}

\begin{figure}
\includegraphics{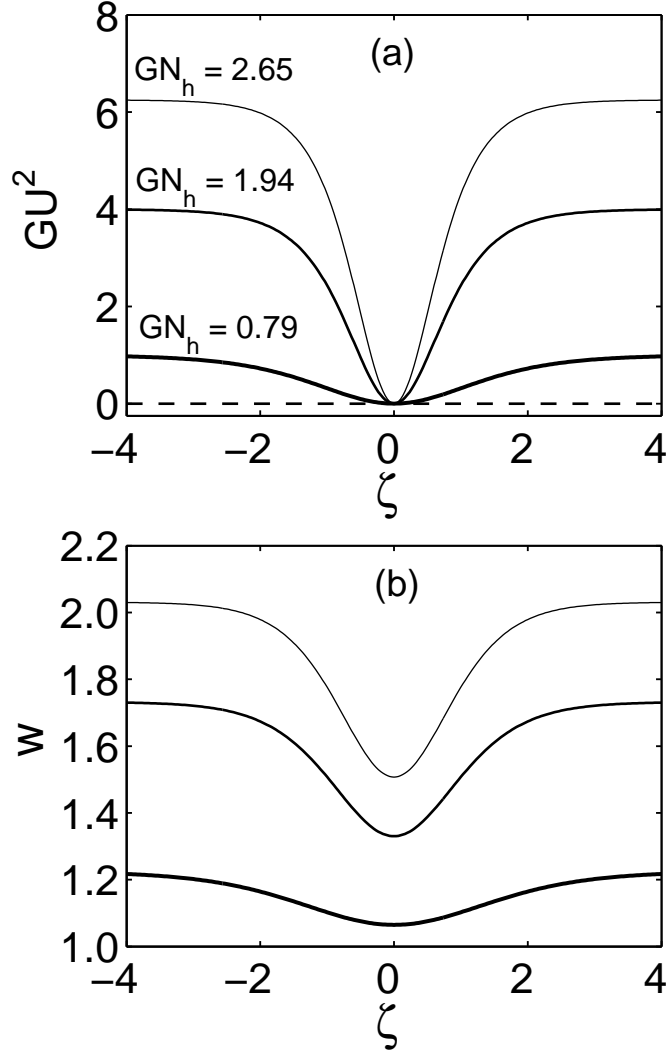}
\caption{\label{F1} The density of the dark soliton solution (a) and its
transverse radius (b) vs. the axial distance $\zeta$ for several values of
the number of atoms in the soliton ``hole''. }
\end{figure}

For large enough $N$ the right inequality in Eq. (\ref{dark-cond2}) is
violated and we have to return to the system (\ref{eq-U2}) and
(\ref{eq-sigma}). Since the left inequality in Eq.~(\ref{dark-cond2}) is
easily fulfilled for a dense BEC in a cigar-shaped trap, we suppose that
the longitudinal size $\sim\zeta_m$ of the whole condensate is much greater
than the soliton width. To find the  soliton solution we take $V(\zeta)=0$
and require  that $U$ satisfies the following asymptotic boundary
conditions:
\begin{equation}\label{bound1}
    |U|\to U_0\quad \text{at}\quad \zeta\to \pm\infty.
\end{equation}
Then Eq.~(\ref{eq-sigma}) (see also Eq.~(\ref{alg-rel})) gives
\begin{equation}\label{dark-sigma}
    \sigma\to\sigma_0=(1+2GU_0^2)^{-1/2}
    \quad \text{as}\quad \zeta\to \pm\infty,
\end{equation}
while from Eq.~(\ref{eq-U2}) (see Eq.~(\ref{mu-stat})) we find the value of
$\mu$:
\begin{equation}\label{mu-sol}
    \mu=\frac{1+3GU_0^2}{\sqrt{1+2GU_0^2}}.
\end{equation} Thus we arrive at the system:
\begin{equation}\label{eq-U2-sol}
    U_{\zeta\zeta}+\left(2\mu-\sigma-\frac1\sigma
    -\frac{\sigma_\zeta^2}{4\sigma^2}\right)U-4G\sigma{U^3}=0,
\end{equation}
\begin{equation}\label{eq-sigma-sol}
    \frac12\left(\frac{\sigma_\zeta U^2}{\sigma}\right)_\zeta
    -\left(\sigma-\frac1\sigma\right)U^2-2G\sigma U^4=0,
\end{equation}
whose solution minimizes the energy
\begin{equation}\label{en-2}
    \mathcal{E}=\int\left[\frac12U_\zeta^2+G\sigma U^4+
    \frac12\left(\sigma+\frac1\sigma+\frac{\sigma_\zeta^2}{4\sigma^2}
    \right)U^2-\mu U^2\right]d\zeta,
\end{equation}
where $\mu$ is fixed by  the asymptotic value $U_0$ of the wave function as
dictated by formula (\ref{mu-sol}).

\begin{figure}
\includegraphics{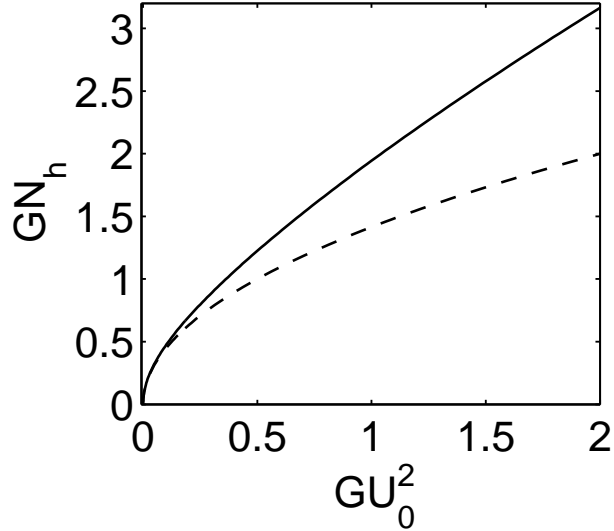}
\caption{\label{F2} The number of atoms in the dark soliton ``hole'' as
function of the asymptotic value $GU^2_0$. The solid line corresponds to
our quasi-1D approach and the dashed line to the NLS approximation.}
\end{figure}

The solution of system (\ref{eq-U2-sol})--(\ref{eq-sigma-sol}) was found
numerically and the results are presented in Fig.~1. As we see, even for
relatively small values of $U_0$ (scaled by $\sqrt{G}$ in the figure)  the
transverse radius of the density distribution changes quite considerably
along the trap axis rather than being constant as is supposed in the NLS
equation approximation. The soliton width is about unity (in the
dimensionless units) for our numerical solutions which also contradicts the
condition (\ref{dark-sol-cond}) of applicability of the NLS equation.

The discrepancy between the improved quasi-1D description and the NLS
solution can be manifested by comparison of the  number of BEC atoms
expelled from the dark soliton ``hole'', defined as
\begin{equation}\label{hole}
    N_{\mathrm{h}} = \int(U_0^2-U^2)d\zeta.
\end{equation}
In Fig.~2 we give the plot of the number of expelled atoms vs. $GU^2_0$. It
shows very convincingly that the NLS approximation is quantitatively
incorrect outside the region $GU^2_0\ll 1$.

We conclude this section by noticing that our quasi-1D theory permits one
to describe dark solitons in  the high density BEC when the standard NLS
approximation breaks down.

\subsection{Bright solitons in BEC with attractive interaction
 ($G<0$)}

In the small density limit (\ref{low-dens-crit}) the condensate with
attractive interaction between the atoms is also described by the NLS
equation (\ref{NLS}), now with $G<0$. In this case BEC can be confined in
the axial direction of the cigar-shaped trap by the interatomic interaction
only. Supposing the latter, we can omit the longitudinal trap potential
$V(\zeta)$ and obtain the bright soliton solution
\begin{equation}\label{bright-sol}
    U(\zeta)=\frac{U_0}{\cosh(U_0\sqrt{2|G|}\,\zeta)},
\end{equation}
where $U_0$ is the amplitude of the soliton connected with the
number of atoms $N=\int U^2d\zeta$ by the relation
\begin{equation}\label{N-sol}
    N=U_0\sqrt{\frac2{|G|}},
\end{equation}
and $\mu'=\mu-1=-|G|U_0^2$. Then  condition (\ref{low-dens-crit}) of
applicability of the NLS equation can be written in the form
\begin{equation}\label{cond-NLS}
    |\mu'|=|G|U_0^2\sim(|G|N)^2\ll1.
\end{equation}
It is clear that this condition breaks down for large enough number of
atoms $N\sim1/|G|$ and one cannot neglect the effect of the atomic
interaction on the  transverse size of BEC. The transverse degrees of
freedom lead to the collapse instability of BEC  (see, e.g. \cite{PS}),
which, as is known, also takes place in BEC confined in a cigar-shaped
trap. Numerical results of Ref.~\cite{Gammal02} indicate that BEC in the
cigar-shaped trap collapses for
\begin{equation}\label{collapse-cond1}
    |G|N>0.676
\end{equation}
and in this region of parameters the soliton solution ceases to exist. Our
quasi-1D approach captures this essential property of the attractive BEC,
and hence, in our approach, the bright soliton solution can be studied in the
whole region of its existence.

\begin{figure}
\includegraphics{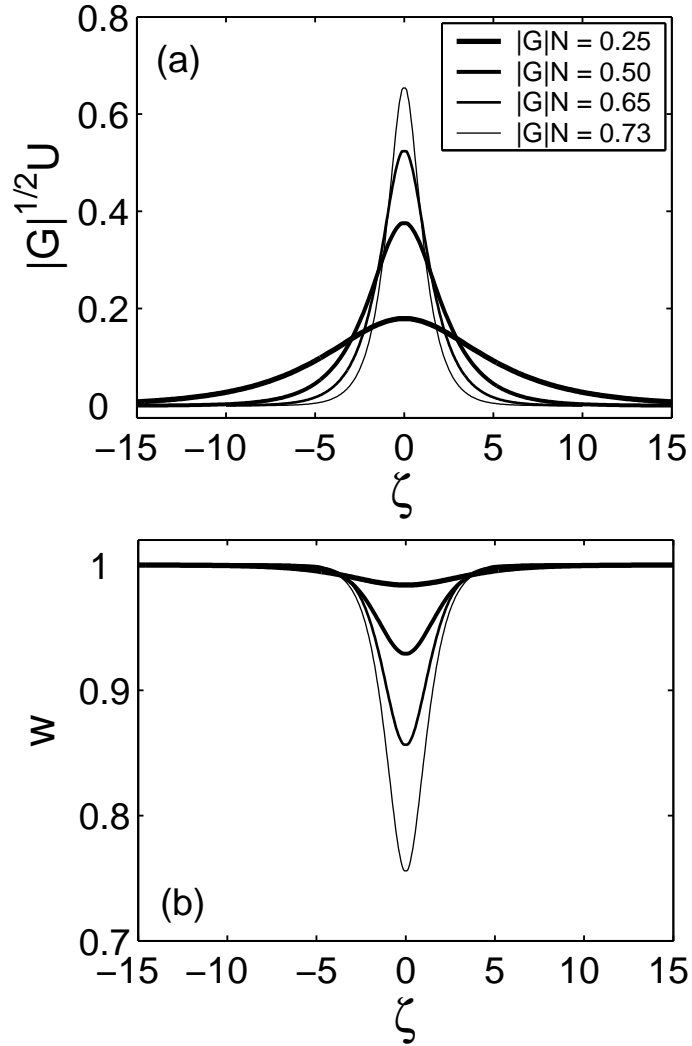}
\caption{\label{F3} The wave function amplitude (a) and transverse radius
(b) of BEC  corresponding to the  bright soliton solution  vs. $\zeta$ for
several values of $|G|N$. }
\end{figure}

We have solved numerically the system (\ref{eq-U2-sol})--(\ref{eq-sigma-sol})
with $G<0$ under the boundary conditions
\begin{equation}\label{bound2}
    U(\zeta)\to 0\quad \text{at}\quad \zeta\to\pm\infty.
\end{equation}
In Fig.~3 we show the wave function amplitude and transverse radius
profiles of BEC for several values of $|G|N$. It is clearly seen that the
amplitude of the soliton  increases while its transverse  radius decreases
with growth of $|G|N$. For small $|G|N$ the solution is well approximated
by the NLS solution (\ref{bright-sol}), especially the amplitude profile.
At $\zeta\to\pm\infty$ the radius approaches $1$, what  corresponds to the
small amplitude limit when the transverse wave function is given by the
oscillator ground state. Fast decrease of $w$ in the center of the soliton
with growth of $|G|N$ precedes the collapse instability of  BEC.

To clarify the transition to collapse, in Fig.~4 we  show the dependence of
$\mu$ on $|G|N$ in our quasi-1D approximation (the solid line) and in the
NLS equation case (the dashed line). We see that $\partial N/\partial \mu$
vanishes at
\begin{equation}\label{collapse-cond2}
    |G|N\cong 0.73
\end{equation}
which is in reasonably good agreement with the exact 3D critical value
given by (\ref{collapse-cond1}).
\begin{figure}
\includegraphics{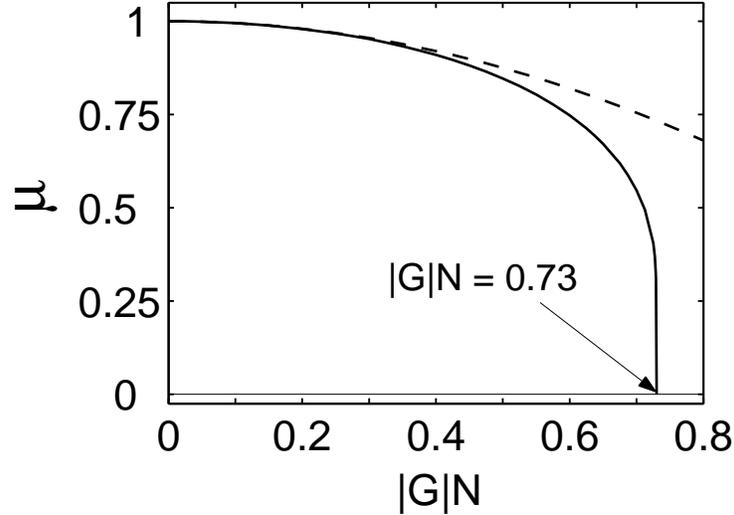}
\caption{\label{F4} The dependence of $\mu$ on $|G|N$ in quasi-1D
approximation (solid line) and in the NLS equation case (dashed line). }
\end{figure}

Finally, the  axial width  of our soliton solution  is about unity (i.e.
$\sim a_\bot$ in the dimensional units) in qualitative agreement with the
features of bright solitons observed experimentally in Refs.
\cite{khaykovich,strecker}.

\section{Linear waves in condensate with repulsive interaction}

Now let us consider the linear waves propagating along a cylindrical
condensate, that is we assume that the wavelength is much less than the
axial size of the condensate so that the influence of the axial trap can be
neglected.

We shall start from equations in the hydrodynamical form
(\ref{cont})--(\ref{cont-2}) with $V'(\zeta)=0$ in Eq.~(\ref{euler}). In
the stationary state the condensate has constant linear density $\rho_0$
and the transverse radius $w_0$  related with $\rho_0$ by the equation (see
(\ref{alg-rel}))
\begin{equation}\label{13-3}
    w_0^4=1+2G\rho_0.
\end{equation}
In the presence of a linear wave $\rho$ and $w$ are slightly deviated from
these stationary values, and a wave can be described by the following
small variables
\begin{equation}\label{14-1}
    \rho'=\rho-\rho_0,\quad w'=w-w_0,\quad v,\quad \beta.
\end{equation}
Linearization of the system (\ref{cont})--(\ref{cont-2}) with respect to these
small variables leads to the following equations:
\begin{equation}\label{14-2}
    \rho'_\tau+\rho_0v_\zeta+\tfrac12\rho_0w_0^2\beta_{\zeta\zeta}=0,
\end{equation}
\begin{equation}\label{14-3}
    v_\tau+\frac{3G}{w_0^2}\rho'_\zeta-2\left(w_0+\frac{G\rho_0}{w_0^3}
    \right)w'_\zeta-\frac1{4\rho_0}\rho'_{\zeta\zeta\zeta}+
    \frac1{2w_0}w'_{\zeta\zeta\zeta}=0,
\end{equation}
\begin{equation}\label{14-5}
    \beta_\tau-\frac{2G}{w_0^4}\rho'+\frac4{w_0}w'-\frac1{w_0^3}
    w'_{\zeta\zeta}  = 0.
\end{equation}
\begin{equation}\label{14-4}
    w'_\tau-w_0\beta+\tfrac14w_0^3\beta_{\zeta\zeta}=0,
\end{equation}

We look for the solution in the form
$(\rho',\,w',\,v,\,\beta)\propto \exp[i(K\zeta-\Omega\tau)]$, the most
interesting limiting cases of which can easily be investigated. Indeed, in
the long wavelength limit $K\ll1$  we can neglect the terms with
higher derivatives. Then Eq.~(\ref{14-4}) gives $\beta=w'_\tau/w_0$ and the
substitution of this into Eq.~(\ref{14-5})  yields the equation
\begin{equation}\label{14-6}
    w'_{\tau\tau}-\frac{2G}{w_0^3}\rho'+4w' =0.
\end{equation}
For low frequency sound wave $\Omega^2\ll4$ and we obtain
$w'=(G/2w_0^3)\rho'$. Then after elimination of the radial variables
$\beta$ and $w'$  the system (\ref{14-2})--(\ref{14-5}) reduces to a couple
of equations for the longitudinal variables:
\begin{equation}\label{14-7}
    \rho'_\tau+\rho_0v_\zeta=0,
    \quad v_\tau + \frac{G}{w_0^2}\cdot\frac{2+3G\rho_0}{1+2G\rho_0}
    \rho'_\zeta = 0.
\end{equation}
These lead at once to the expression for the sound velocity
\begin{equation}\label{14-8}
    c_s=\frac{\Omega}{K}, \quad  c_s^2=\frac{G\rho_0(2+3G\rho_0)}{(1+2G\rho_0)^{3/2}}.
    \end{equation}

In dimensional units we have
\begin{equation}\label{15-1}
    G\rho_0=a_sw_0^2n_1
\end{equation}
where
\begin{equation}\label{14-9}
    n_1=\pi a_\bot^2n
\end{equation}
is the one-dimensional density in the limit of tight radial trapping
($G\rho_0\ll1$) and $n$ is the three-dimensional density evaluated in the
``center of the trap'' at $r=0$. Thus $w_0$ is obtained from the equation
$w_0^4=1+2aw_0^2n_1$ which yields
\begin{equation}\label{14-12}
    w_0=\left[\sqrt{1+(an_1)^2}+an_1\right]^{1/2}.
\end{equation}
Then, in the dimensional units, we obtain the following expression for the
velocity of sound  propagating along the trap:
\begin{equation}\label{14-10}
    \tilde{c}_s^2=\left(\frac{\hbar}{ma_\bot}\right)^2an_1\cdot\frac{2+3G\rho_0}
    {1+G\rho_0}.
\end{equation}
In the low density (tight radial trapping) limit we have $G\rho_0\sim a_sn_1\ll1$,
so that formula (\ref{14-10}) reproduces the well-known result (see, e.g.
Refs.~\cite{PS,zaremba})
\begin{equation}\label{14-11}
    \tilde{c}_s=\frac{\hbar}{ma_\bot}\sqrt{2an_1}.
\end{equation}
Expression (\ref{14-10}) gives dependence of the sound velocity on the
density $n_1$ correct up to  $G\rho_0\sim1$.

Now, if we neglect all $\zeta$-derivatives in Eqs.~(\ref{14-2})--(\ref{14-5}),
then we obtain $\rho'=0,$ $v=0$, i.e. there is
no motion of the gas along the trap, and Eq.~(\ref{14-6}) leads to the
frequency
\begin{equation}\label{14-13}
    \Omega=2
\end{equation}
of compressional oscillations in  radial direction. This formula agrees
with the general statement \cite{PS,kagan96,pit96} that radial oscillations
of a very elongated condensate has the frequency $\Omega=2\omega_\bot$ (in
dimensional units) independent of interatomic interaction.

Equations (\ref{14-2})--(\ref{14-5}) permit us to find corrections to the
frequencies in equations (\ref{14-8}) and (\ref{14-13}). We have the
following dispersion relation for the sound waves
\begin{equation}\label{16-1}
    \Omega^2=c_s^2K^2+\frac14\left[1+\frac{(G\rho_0)^2(1-3(G\rho_0)^2)}
    {(1+2G\rho_0)^3}\right]K^4.
\end{equation}
It is worth noticing that the dispersion correction changes its sign at
$G\rho_0\cong 3.91$ and higher order dispersion correction has to be taken into
account at this point.

For high-frequency radial compression wave we obtain
\begin{equation}\label{16-2}
    \Omega^2=4+\frac{2+6G\rho_0+5(G\rho_0)^2}{(1+2G\rho_0)^{3/2}}K^2.
\end{equation}
Naturally, a correction term is not universal and depends on the atomic
interaction.

\section{Weakly dispersive and nonlinear waves in repulsive BEC}

At last, let us turn to waves of small but finite amplitude which propagate
in repulsive BEC  confined in an elongated cigar-shaped trap.

As we know from the preceding section, the linear waves of long wavelength
propagate with the sound velocity $c_s$. For finite wavelength we
have to take into account the dispersion correction described by the second
term in the dispersion relation (\ref{16-1}). We must also take into
account small correction arising due to nonlinear terms in the system
(\ref{cont})--(\ref{cont-2}) (with $V'(\zeta)=0$). To this end, we use the
 standard singular perturbation theory (see, e.g., \cite{kamch}) and
introduce the stretched variables
\begin{equation}\label{73}
    T=\eps^{3/2}\tau,\quad Z=\eps^{1/2}(\zeta+c_s\tau)
\end{equation}
corresponding to the slow evolution of the wave in the reference system
moving with the sound velocity $c_s$. Here $\eps$ is a small parameter
which characterizes the magnitude of a perturbation of the ground state due
to presence of wave. Hence, we introduce the series expansions in powers of
$\eps$:
\begin{equation}\label{74}
    \begin{split}
    &\rho =  \rho_0+\eps\rho^{(1)}+\eps^2\rho^{(2)}+\ldots,\quad
    w=w_0+\eps w^{(1)}+\eps^2 w^{(2)}+\ldots,\\
    &v =  \eps v^{(1)}+\eps^2 v^{(2)}+\ldots,\quad
    \beta=\eps^{3/2}\beta^{(1)}+\beta^{5/2}\beta^{(2)}+\ldots.
    \end{split}
\end{equation}
Substitution of these series expansions into Eqs.~(\ref{cont})--(\ref{cont-2})
gives, in the first two orders in powers of $\eps$ the ground state
formula (\ref{13-3}), the relationships
\begin{equation}\label{75}
    w^{(1)}=\frac{G}{2w_0^3}\rho^{(1)},\quad
    v^{(1)}=-\frac{c_s}{\rho_0}\rho^{(1)},\quad
    \beta^{(1)}=\frac{Gc_s}{2w_0^4}\rho^{(1)}_Z,
\end{equation}
and reproduces the formula (\ref{14-8}) for the sound velocity.
The last relevant terms of the expansions in powers of $\eps$
yield the equations
\begin{equation}\label{76}
    -c_s\beta^{(1)}_{ZZ}+\frac1{w_0^3}w^{(1)}_{ZZZ}+
    \frac{20}{w_0^2}w^{(1)}w^{(1)}_Z-\frac{8G}{w_0^5}(\rho^{(1)}w^{(1)})_Z
    =\frac4{w_0}w^{(2)}_Z-\frac{2G}{w_0^4}\rho^{(2)}_Z,
\end{equation}
\begin{equation}\label{77}
    \rho^{(1)}_T+(v^{(1)}\rho^{(1)})_Z+\frac12\rho_0w_0^2\beta^{(1)}_{ZZ}
    =-c_s\rho^{(2)}_Z-\rho_0v^{(2)}_Z,
\end{equation}
\begin{equation}\label{78}
    \begin{split}
    v^{(1)}_T+v^{(1)}v^{(1)}_Z+\frac{6(3G\rho_0+1)}{w_0^4}w^{(1)}w^{(1)}_Z-
    \frac{6G}{w_0^3}(\rho^{(1)}w^{(1)})_Z-\frac1{4\rho_0}\rho^{(1)}_{ZZZ}
    +\frac1{2w_0}w^{(1)}_{ZZZ} \\=
    -c_sv^{(2)}_Z+\frac{2(3G\rho_0+1)}{w_0^3}w^{(2)}_Z-\frac{3G}{w_0^2}\rho^{(2)}_Z,
    \end{split}
\end{equation}
which can be considered as a linear system for the variables
$\rho^{(2)}_Z$, $w^{(2)}_Z$, $v^{(2)}_Z$ (we have omitted the equation
following from Eq.~(\ref{cont-2}) which gives the expression for
$\beta^{(2)}$ in terms of the other variables). It is easy to find that the
right-hand sides of Eqs.~(\ref{76})-(\ref{78}) are linearly dependent.
Hence, for existence of a non-trivial solution, the left-hand sides must
satisfy the corresponding compatibility condition which with account of
Eqs.~(\ref{75}) can be transformed to the KdV equation taking after transition
to the ``laboratory'' reference frame with variables $\tau$, $\zeta$ and the
density disturbance $\rho'=\eps\rho^{(1)}$ the form:
\begin{equation}\label{79}
    \rho'_\tau-c_s\rho'_\zeta+\frac1{8c_s}\left[1+\frac{(G\rho_0)^2(1-3(G\rho_0)^2)}
    {(1+2G\rho_0)^3}\right]\rho'_{\zeta\zeta\zeta}-\frac{3G}{c_sw_0^2}
    \left[1+\frac{(G\rho_0)^2}{2(1+2G\rho_0)^2}\right]\rho'\rho'_\zeta=0.
\end{equation}
The linear terms here reproduce the first two terms (\ref{16-1}) of the
expansion of the dispersion relation in powers of $K$. This KdV equation has
well-known soliton solutions.

In the low density limit $G\rho_0\ll1$  Eq.~(\ref{79}) takes the form
\begin{equation}\label{80}
    \rho'_\tau-c_s\rho'_\zeta+\frac1{8c_s}\rho'_{\zeta\zeta\zeta}-
    \frac{3G}{c_s}\rho'\rho'_\zeta=0,\quad G\rho_0\ll1.
\end{equation}
It is worth while to note that this equation can be derived directly from
the 1D NLS equation (\ref{c2}) which is a correct approximation to the GP
equation in  this limit. Indeed, evolution of a small-amplitude disturbance
$a(z,t)$ of the ground state $\Psi_0=\mathrm{const}$, i.e.
$\Psi=[\Psi_0+a(z,t)]\exp(-i\om t+i\phi(z,t))$ is governed by the KdV
equation (see, e.g. \cite{kivshar})
\begin{equation}\label{81}
    a_t-\tilde{c}_sa_z+\frac{\hbar^2}{8m^2\tilde{c}_s}a_{zzz}-
    \frac{3g_{1D}\Psi_0}{m\tilde{c}_s}aa_z=0,
\end{equation}
where $\tilde{c}_s$ is the dimensional sound velocity (\ref{14-11}). Then
transformation of this equation to dimensionless units (\ref{dim}) with
account of the relationship $\rho'\cong2a_\bot\Psi_0 a$ between small
disturbances of the density ($\rho'$) and the amplitude ($a$) reproduces
Eq.~(\ref{80}).

A remarkable feature of Eq.~(\ref{79}) is the fact that the dispersion
term changes sign at $G\rho_0=3.91$ while the coefficient at the nonlinear
term is always negative. This means that the small amplitude solitons
propagating on a constant background in repulsive BEC  are dark only in the
region $G\rho_0<3.91$ but become bright in the region $G\rho_0>3.91$.
Obviously, such behavior is explained physically by competition of
two mechanisms of energy changes due to presence of the wave---one is
caused by the change of the potential energy of the gas in the transverse
trap and another one is related with the nonlinear energy of interatomic
gas interaction.
Thus, we have found that bright solitons can propagate along dense enough
repulsive BEC confined in a cigar-shaped trap. These solitons propagate
with the velocity close to the speed of sound.

\section{Conclusion}

The effective one-dimensional system derived in this paper describes motion
of condensate in a wide region of parameters when the radial distribution
has not reached yet the Thomas-Fermi limit and the characteristic axial
dimension is not less than the BEC radius in an elongated cigar-shaped
trap. This system depends  only on one spatial coordinate, what can greatly
simplify numerical simulations of nonlinear effects in BEC. The advantage
of our approach is illustrated by applications to description of dark and
bright solitons in the high density BEC. The one-dimensionality of the
effective system permits one to develop various analytical approximations.
In particular, the dispersion relations for sound and radial compression
waves are obtained for high density BEC. Evolution of small but finite
amplitude perturbations is described by the KdV equation which leads to
qualitatively new behavior of BEC at high enough density: small-amplitude
bright  solitons can propagate on a constant background in a repulsive BEC,
whose density is above some threshold value. This prediction can be
verified by experiment.

We hope that the quasi-1D theory developed in this paper will find
applications in description of many other nonlinear processes in the
high-density BEC confined in the cigar-shaped trap, i.e. when the standard
one-dimensional nonlinear Schr\"odinger equation looses its applicability.

\section*{Acknowledgements}

This work was supported by the FAPESP and FAPEAL/CNPq grants (Brazil).
A.M.K. thanks also RFBR for partial support.

\end{document}